# Curriculum Guidelines for Undergraduate Programs in Data Science

**Park City Math Institute (PCMI) Undergraduate Faculty Group***

Richard De Veaux (chair), Mahesh Agarwal, Maia Averett, Benjamin Baumer, Andrew Bray, Thomas Bressoud, Lance Bryant, Lei Cheng, Amanda Francis, Robert Gould, Albert Y. Kim, Matt Kretchmar, Qin Lu, Ann Moskol, Deborah Nolan, Roberto Pelayo, Sean Raleigh, Ricky J. Sethi, Mutiara Sondjaja, Neelesh Tiruviluamala, Paul Uhlig, Talitha Washington, Curtis Wesley, David White, Ping Ye

*Author affiliations can be found in the Acknowledgments section.

**Abstract**

The Park City Math Institute (PCMI) 2016 Summer Undergraduate Faculty Program met for the purpose of composing guidelines for undergraduate programs in Data Science. The group consisted of 25 undergraduate faculty from a variety of institutions in the U.S., primarily from the disciplines of mathematics, statistics and computer science. These guidelines are meant to provide some structure for institutions planning for or revising a major in Data Science.

# 1. Introduction

Data Science is experiencing rapid and unplanned growth, spurred by the proliferation of complex and rich data in science, industry and government. Fueled in part by reports such as the widely cited McKinsey report that forecast a need for hundreds of thousands of Data Science jobs in the next decade (McKinsey), Data Science programs have exploded in academics as university administrators have rushed to meet the demand. The website *http://datascience.community/colleges* currently lists 530 programs in Data Science, analytics and related fields at over 200 universities around the world. The vast majority of these are master's degree and certificate programs offered both traditionally and online. While PhD programs in Data Science (or Data Analytics) are still relatively rare, there has been rapid growth of undergraduate programs at both research institutions and liberal arts colleges. We expect this number to increase significantly in the near future.

The 2016 Park City Mathematics Institute (PCMI), sponsored by the National Science Foundation (NSF) and the Institute for Advanced Study at Princeton (IAS), held a workshop for undergraduate faculty focused on the task of producing curriculum guidelines for an undergraduate degree in Data Science. Twenty-five faculty, comprised of computer scientists, statisticians and mathematicians from a variety of liberal arts colleges and research universities, met for three weeks to discuss our vision for Data Science in an undergraduate context, what activities and skills we thought would be necessary for a Data Science program and how we could imagine implementing such a major both currently and in the future. These guidelines are the product of that effort.

We have based our guidelines for an undergraduate Data Science major on a ten semester-course major common among the liberal arts colleges, realizing that research universities typically add several courses to that. We do not intend that these guidelines be prescriptive, but rather we hope that they will serve to inform and enumerate the core skills that a Data Science major should have before graduation. We started with the reports from the NSF Workshop on Data Science Education (See *"Strengthening Data Science Education through Collaboration"*, October 1-3, 2015), the AALAC Big Data Conference (Wellesley, January 2016) and the guidelines for undergraduate majors in Mathematics, Statistics and Computer Science (see sidebar).

---

**Curriculum Guidelines in Related Disciplines**

**2015 CUPM Curriculum Guide to Majors in the Mathematical Sciences (MAA)**
*http://www.maa.org/sites/default/files/pdf/CUPM/pdf/CUPMguide_print.pdf*
**Computer Science Curricula 2013: Curriculum Guidelines for Undergraduate Degree Programs in Computer Science (ACM)** *https://www.acm.org/education/CS2013-final-report.pdf*
**Curriculum Guidelines for Undergraduate Programs in Statistical Science (ASA)**
*http://www.amstat.org/education/pdfs/guidelines2014-11-15.pdf*

---

We begin by discussing the background and some guiding principles that informed our thinking in Section 2, then consider skills that students should develop while pursuing the major in Section 3, and finally summarize key curriculum topics in Sections 4 and 5. We show a possible selection of current courses that do cover most of the basics of our identified skills in Section 6. However, it is important to point out that this smorgasbord approach to course selection is less than ideal. We believe that many of the courses traditionally found in computer science, statistics and mathematics offerings should be redesigned for the Data Science major for the sake of efficiency and of the potential synergy that integrated courses would offer. Relying on existing courses at most institutions, a student might have to take 14 or more courses in order to obtain all the skills one would expect from a Data Science major. With some significant course redesign, we think that this number could be substantially reduced to fit into the constraints of a typical 10-course liberal arts major. Details of those courses are found in the Appendix.

## 2. Background and Guiding Principles

Our curriculum guidelines are guided both by recent work on the relation of data science to the other sciences as well as the need for a workforce better able to meet the demands of the data-driven economy of the future.

### 2.1. Data Science as Science

Even though an exact definition of Data Science remains elusive, we have taken as our starting point a view that seems to have emerged as a consensus from the StatSNSF committee statement that Data Science comprises the "science of planning for, acquisition, management, analysis of, and inference from data." See *http://www.nsf.gov/attachments/130849/public/Stodden-StatsNSF.pdf*. At the undergraduate level, we conceive of Data Science as an applied field akin to engineering, with its emphasis on data and how it describes the world. At present, the theoretical foundations are drawn primarily from established strains in Statistics, Computer Science, and Mathematics. The practical real-world meanings come from interpreting the data in the context of the domain in which the data arose. For an undergraduate program, we envision a case-based focus and hands-on approach, as is common in fields such as engineering and computer science.

### Is Data Science a Science?

There is still considerable debate about exactly what the science of Data Science is, but prominent scientists such as David Donoho, Michael Jordan and others suggest that there is a science at the core and that it will continue to evolve. As Donoho says, "Fortunately, there is a solid case for some entity called 'Data Science' to be created, which would be a true science: facing essential questions of a lasting nature and using scientifically rigorous techniques to attack those questions." Regardless of the consensus (or lack thereof) surrounding the evolution of the science of Data Science, a Data Science program at the undergraduate level provides a synergistic approach to problem solving, one that leverages the content in all three disciplines. We believe that a Data Science program will serve students well whether they join the marketplace or continue on to more advanced study.

## 2.2. Interdisciplinary Nature of Data Science

Data science is inherently interdisciplinary. Working with data requires the mastery of a variety of skills and concepts, including many traditionally associated with the fields of Statistics, Computer Science and Mathematics. Data Science blends much of the pedagogical content from all three disciplines, but it is neither the simple intersection, nor the superset of the three. By applying the concepts needed from each discipline in the context of data, the curriculum can be both significantly streamlined and enhanced. The integration of courses, focused on data, is a fundamental feature of an effective Data Science program and results in a synergistic approach to problem solving.

This document outlines the core knowledge and methods that Data Science students should master. Our position is that, ideally, new courses should be developed to take advantage of the efficiencies and synergies that an integrated approach to Data Science would provide. However, since not all institutions will be able to create many new courses immediately, we suggest which traditional courses might provide coverage of the basic topics of the major. We also propose a model of an integrated curriculum to serve as a possible blueprint for the future.

## 2.3. Data at the Core

The recursive data cycle of obtaining, wrangling, curating, managing and processing data, exploring data, defining questions, performing analyses and communicating the results lies at the core of the Data Science experience. Undergraduates need understanding of, and practice in performing all steps of this data cycle in order to engage in substantive research questions. In the words of Google's Diane Lambert students need the ability to "think with data" (Horton & Hardin) (See also (ASA) and (Shron 2014)). Data *experiences* need to play a central role in all courses from the introductory course to the advanced elective/capstone. These experiences should include raw data from a variety of sources and include the process of cleaning, transforming, and structuring data for analysis. They should also include the topic of data provenance and how that informs the conclusions one can draw from data. Data science is necessarily highly experiential; it is a practiced art and a developed skill. Students of Data Science must encounter frequent project-based, real-world applications with real data to complement the foundational algorithms and models. The Committee on the Undergraduate Program in Mathematics Curriculum Guide from 2004 and 2015 reinforced the importance of real applications and data analysis (MAA) (MAA 2015) for all mathematical science majors. They stated that:

> the analysis of data provides an opportunity for students to gain experience with the interplay between abstraction and context that is critical for the mathematical sciences major to master. Experience with data analysis is particularly important for majors entering the workforce directly after graduation, for students with interests in allied disciplines, and for students preparing to teach secondary mathematics.

Statisticians, naturally, feel the same. In 2001, a report by the ASA recommended that undergraduate statistics curriculum include "a heavy emphasis on data analysis (perhaps more weight should be given to the data than the analysis)" (ASA 2015). By the same logic, students learn Data Science by doing Data Science. The recursive data cycle should be a featured component of most Data Science learning experiences and projects involving group analysis and presentation should be common throughout the curriculum. Capstone projects are also an essential component of the experience and internships fit naturally in a Data Science program.

## 2.4. Analytical (Computational and Statistical)Thinking

Breiman (2001) spoke of the two cultures of Algorithmic (computational) and Data (statistical) models (renamed predictive and inferential by Donoho *opcit*). Data science offers the opportunity to integrate and use both computational and statistical thinking to problem solving rather than emphasizing one over the other. Interestingly, these are not new ideas. As Wilkinson (2008) pointed out: 'Many of Tukey's ideas are now conventional wisdom. For example, Tukey accorded algorithmic models the same foundational status as the algebraic (data) models that statisticians had favored in the previous half-century'. The two pillars of computational and statistical thinking should not be taught separately. The balance between them may change from one course to another, but both should be present for the most effective and efficient teaching.

## 2.5. Mathematical Foundations

Data scientists employ models to understand the world and mathematics provides the language for these models, so a working data scientist requires a firm foundation in mathematics. However, traditional mathematics curricula often delay the connection between abstract mathematics and messy, possibly ill-posed, real world problems, especially with respect to those involving data. We propose a fresh approach that distills the essential aspects of mathematics needed for Data Science at the undergraduate level. Rather than requiring four or more courses in mathematics, an efficient Data Science major should present these mathematical concepts in two courses, in the context of modeling for data-driven problems. This will streamline the mathematical curriculum to focus on Data Science rather than theory, derivations, or proofs. In particular, we propose modeling (both algorithmic and statistical) as a motivator for mathematical tool development, introducing concepts as they become necessary in order to solve our real-world problems. Matrix algebra is motivated by solving linear systems, derivatives are motivated by optimization and sensitivity analysis, integration is motivated by probabilistic applications. While this shift towards modeling and applications is in line with current trends in calculus reform, our approach is revolutionary in its approach to the mathematics, in its ordering of topics, and in the selection of topics.

A possible caveat is that our approach is not meant to serve as an alternative pathway for the mathematics major, though students introduced to mathematics this way may want to further their study. Such students would need additional theoretical foundations before entering traditional upper-division mathematics courses.

## 2.6. Flexibility

We must prepare students to learn new techniques and methods that may not exist today. They will need to work with increasingly varied forms of data, or they will not be prepared for the jobs of the future. We need to pay attention to the core foundations of mathematics, computational and statistical thinking and practice while incorporating the practical and important Data Science skills.

Data Science, at all levels, is evolving and changing quickly. Most institutions will implement a Data Science major from current courses in existing disciplines, perhaps transitioning to more fully integrated courses as outlined in the Appendix at a future date. Our hope is that institutions use these guidelines in their planning to meet the needs of their students both now and in the future. We fully expect that institutions will regularly review their programs to reflect new developments in this fast-evolving field.

## 3. Key Competencies and Features of a Data Science Major

A graduate with an undergraduate Data Science degree should be prepared to interact with data at all stages of an investigation and will be expected to work within a team environment.

> **Key Competencies for an undergraduate Data Science Major**
>
> **Computational and Statistical Thinking**
> **Mathematical Foundations**
> **Model Building and Assessment**
> **Algorithms and Software Foundation**
> **Data Curation**
> **Knowledge Transference – Communication and Responsibility**

### 3.1. Analytical (Computational and Statistical) Thinking

Data science consists of a problem-solving approach for working within empirical settings in which meaning must be extracted from data. This approach is a synthesis of modes of thought in statistics, computer science, and mathematics.

1. **Statistical Thinking in a Data-Rich Environment**
   Statistical thinking is an approach to understanding the world through data, and involves everything "from problem formulation through conclusions" (Wild & Pfannkuch). The data scientist needs an understanding of basic statistical theory. Students should understand the basic statistical concepts of data analysis, data collection, modeling, and inference. A sound knowledge of basic theoretical foundations will help inform their analyses and the limits to their models. Successful graduates will be able to apply statistical understandings and computational skills to formulate problems, plan data collection campaigns or identify and gather relevant existing data, and then analyze the data to provide insights.
2. **Computational Thinking**
   Working with data requires extensive computing skills. A Data Science student must be prepared to work with data as they are commonly found in the workplace and research labs. For example, accessing and organizing data in databases, scraping data from websites, processing text into data that can be analyze and ensuring secure and confidential data storage all require extensive computing skills. These computational problem-solving skills recur throughout the workflow of the data scientist. As Jeannette Wing put it, "Thinking like a computer scientist means more than being able to program a computer. It requires thinking at multiple levels of abstraction" (Wing). Data Science graduates should be proficient in many of the foundational software skills and the associated algorithmic, computational problem-solving of the discipline of computer science. To be prepared for careers in Data Science, students also need facility with, and exposure to professional statistical analysis software packages, and an understanding of the underlying principles of programming and algorithmic problem-solving that underlie these packages.

3. **Integration of Approaches**

    Data Science at the undergraduate level combines computational and statistical thinking practices, relying on mathematical foundations. In addition to their competence in the areas noted below, Data Science graduates should demonstrate an understanding of the connections between these knowledge domains. They should be able to bring a wide array of different skills and problem-solving approaches to bear on any particular problem, and should make informed choices about which skills are appropriate in a given setting. They should have facility for working with a diverse collection of tools, as well as learning new tools and—in some cases—contributing to the development of new tools themselves.

    Computing environments, both software and hardware, change rapidly and frequently, and these changes have consequences for data structure, data storage, and computational efficiency. Data scientists must be capable of adapting smoothly to such changes. Data scientists should understand both the computational and modeling challenges in their work, and how they might be intertwined. For example, data scientists should recognize that fitting a given model on a particular set of data will engender computational challenges, and they should have some facility for implementing a solution that may involve either a modification of the model or a change in the computing environment, or both. When integrated with statistical thinking, computational thinking greatly amplifies the ability of data scientists to distribute solutions to clients, to understand many modern statistical modeling approaches, and to achieve scientific reproducibility.

## 3.2. Mathematical Foundations

Mathematically speaking, the emphasis of an undergraduate Data Science degree should be on choosing, fitting and using mathematical models. Because data-driven problems are often messy and imprecise, students should be able to impose mathematical structure on these problems by developing structured mathematical problem solving skills. Students should have enough mathematics to understand the underlying structure of common models used in statistical and machine learning as well as the issues of optimization and convergence of the associated algorithms. Although the tools needed for these include calculus, linear algebra, probability theory and discrete mathematics, we envision a substantial realignment of the topics within these courses and a corresponding reduction in the time students will spend to acquire them.

## 3.3. Model Building and Assessment

1. **Informal Modeling**

Statistical models are used to describe, predict, and explain processes, but they are also used to communicate understandings and to lay foundations for future models. Informal modeling involves identifying potential sources of variation, discerning between stochastic and deterministic variation, and understanding how these might be modeled mathematically and computationally. Graduates must also be adept at data visualization, which is an important tool in informal modeling as it can communicate with others and identify weaknesses in proposed models.

2. **Formal Modeling**
   Graduates should be able to build and assess statistical and machine learning models, employ a variety of formal inference procedures, and draw appropriate scope of conclusions from the analysis. This includes understanding how data issues (*e.g.*, collection methods, sources of bias and variance) impact the analysis and interpretation and generalization of statistical findings. Graduates should also be able to bring computational considerations to bear in the analysis of data, including issues of scale.

### 3.4. Algorithms and Software Foundation

The Data Science graduate should be able to employ algorithmic problem solving skills to the task at hand. These include: defining clear requirements to a problem, decomposing the problem, using efficient strategies to arrive at an algorithmic solution and implementing solutions through programming in a suitable high level language. They should understand the memory and execution performance of the structures and software they create, and that of the libraries and packages they use. They should know and utilize good practices in documentation and structure and be able to use appropriate tools for maintaining their software. They should be able to leverage existing packages and tools to solve their computational problems.

### 3.5. Data Curation

Data curation involves managing data through the entire problem-solving process.

1. **Data Preparation**
   Graduates should be able to work with data from a variety of sources and formats. Data may come from a web page, a database, or a stream, and may consist of images, sounds, video as well as numbers or text. These data may have been collected through a controlled experiment, an observational study, or may be opportunistic data collected through sensors or an automated procedure. Given a particular data set, graduates should be able to prepare the data for use with a variety of statistical methods and models and should recognize how the quality of the data and the means of data collection may affect conclusions.
2. **Data Management**
   Data scientists must do more than prepare data for analyses, but must ensure the integrity of the data while it passes through all stages of the analysis. This requires working with relational databases (such as a SQL database), maintaining version control and tracking data provenance as data from multiple sources are merged.

### 3.6. Knowledge Transference

Data do not exist in a vacuum, but arise from a particular context. Knowledge of that context is necessary to analyze the data, and thus undergraduates need experience applying their discipline outside the core of statistics, computing, and mathematics. A Data Science program should feature data-based investigations in a complementary discipline, such as a physical science, a life science, business, a social science, the humanities or fine arts.

1. **Communication**
   Effective communication is a core skill of the data scientist. As a member of a team, data scientists must communicate to teammates as well as to those with less intimate knowledge of the project particulars. Increasingly, data scientists communicate directly to the public via both static and interactive data visualizations. A thoughtful Data Science program integrates communication-based opportunities and learning development throughout the whole of the curriculum rather than partitioning them into separate classes. Students should gain experience using oral, written and visual modes to communicate effectively to a variety of audiences.
2. **Ethics and Reproducibility**
   The capabilities of Data Science introduce new ethical questions. Programs in Data Science should feature exposure to and ethical training in areas such as citation and data ownership, security and sensitivity of data, consequences and privacy concerns of data analysis, and the professionalism of transparency and reproducibility.

### 4. Curricular Content for Data Science Majors

The goal of our curriculum is to repeatedly engage students in the *full cycle* by which we learn from data and to help them acquire the skills listed in the previous section. This necessitates the interweaving and integrating of traditionally siloed topics and tools into a cohesive presentation. Our Data Science program includes six main subject areas, which are described below. We then describe suggested courses that prepare students in all six areas. In the Appendix we show in more detail how courses in each of these cycles might be constructed.

---

**Six Main Subject Areas of a Data Science Major**

**Data Description and Curation**
**Mathematical Foundations**
**Computational Thinking**
**Statistical Thinking**
**Data modeling**
**Communication, Reproducibility and Ethics**

---

A summary of the courses designed for these subject areas is found in the following:

> **An Outline of the Data Science Major**
>
> 1. Intro to Data Science
>    - Intro to Data Science I
>    - Intro to Data Science II
> 2. Mathematical Foundations
>    - Mathematics for Data Science I
>    - Mathematics for Data Science II
> 3. Computational Thinking
>    - Algorithms and Software Concepts
>    - Databases and Data Management
> 4. Statistical Thinking
>    - Intro to Statistical Models
>    - Statistical and Machine Learning
> 5. Course in an Outside Discipline
> 6. Capstone Course

### 4.1. Overview of Course Sequence

- **Intro to Data Science I and II**
  Students will immediately use a high level language to explore, visualize and pose questions about data. In the second semester, a more algorithmic language may be introduced to help students understand the thinking and structure behind the higher level functions they experienced in the first semester.
    - Introduction to high level language
    - Exploring and manipulating data
    - Functions and basic coding
    - Introduction to modeling, both deterministic and stochastic
    - Concepts of projects and code management
    - Databases
    - Introduction to data collection and statistical inference

- **Mathematical Foundations I and II**
  Data science students connect mathematical tools to real-world problems. Unlike pure mathematics, which seeks to build theory and prove propositions, Data Science is about seeing the value of mathematical methods while understanding their limitations. Data science students should also develop a geometric, intuitive, visual way of thinking throughout their mathematical training. We propose a two-semester 'Math for Data Science' sequence that begins with students who would have placed into Calculus 1. This sequence emphasizes mathematical modeling, especially linear and polynomial models. (See the Appendix.)

- Mathematical structures (*e.g.*, functions, sets, relations, logic)
- Linear modeling and matrix computation (*e.g.*, matrix algebra and factorization, eigenvalues/eigenvectors, projection/least-squares)
- Optimization (*e.g.*, calculus concepts related to differentiation)
- Multivariate thinking (*e.g.*, concepts and numerical computation of multivariate derivatives and integrals)
- Probabilistic thinking and modeling (*e.g.*, counting principles, univariate and multivariate distributions, independence, relying often on computational simulations)

- **Model Building and Assessment**
  - Exploratory data analysis approaches and graphical data analysis methods
  - Estimation and testing: Exposure to statistical (*e.g.*, basic central limit theory and law of large numbers) and algorithmic (*e.g.*, bootstrap, resampling methods) approaches to point and interval estimation and hypothesis testing; likelihood theory; and Bayesian methods.
  - Data Transformation: Smoothing and aggregating (*e.g.*, kernel methods, and spatial/temporal smoothing); transforming data for appropriate application of models. Dimension reduction, feature selection, creation and extraction.
  - Simulation and resampling: Monte Carlo simulation of stochastic systems; resampling-based inference (*e.g.*, bootstrap, jackknife, permutations); basic understanding of design of studies, surveys, experiments (*e.g.*, random assignment, random selection, data collection, and efficiency) and issues of bias, causality, confounding, and coincidence
  - Models: Analytic (*e.g.* simple linear, multivariate, and generalized linear models) and algorithmic models (*e.g.*, regression trees); supervised (*e.g.*, nearest neighbors, logistic regression) and unsupervised learning (*e.g.*, clustering). Model averaging and ensemble methods (*e.g.*, boosting and bagging)
  - Model selection and performance: Regularization, parsimony and bias/variance tradeoff; loss functions and model selection (*e.g.* cross-validation, penalized regression, ridge regression)

- **Algorithms and Software Foundation**
  To develop a grounded computational ability, a Data Science undergraduate should study foundational computer science topics and build facility in algorithmic problem solving and development of software/programming.
  - Algorithm design: students must develop the skill set to understand the problem, break it into manageable pieces, assess alternative problem solving strategies and arrive at an algorithm that efficiently solves the problem.
  - Programming concepts and data structures: students should have the knowledge to implement their algorithms using procedural and functional programming techniques and their associated data structures, including lists, vectors, data frames, dictionaries, trees, and graphs.

- Tools and Environments: Students should understand the appropriate use of tools and packages available. Such packages enable programmatic access to data services and I/O, perform data transformations, explorations, visualization, and analysis, and assist in the development and maintenance of software, including development environments, and tools for versioning and tracking.
- Scaling for big data: As the data and processing associated with Data Science continue to scale, Data Science undergraduates should develop the capacity to work with larger datasets. They should be able to apply techniques in concurrent programming to build systems that perform parallel processing of data. They must also be able to work with current and new forms of distributed data storage, as a part of the data management areas discussed above. They should be knowledgeable in how to work with streaming data.

- **Data Curation – Databases and Data Management** A Data Science undergraduate major must understand and be able to effectively apply principles of data management. This is much broader than traditional database management and must include systems supporting the volume and velocity attributed to big data. Thus a Data Science major must apply knowledge of data query languages to relational databases and emerging large store NoSQL data systems, must be able to access data from less structured systems through web services and lower level access to data available across the Internet, and data sourced from streams. Once data are collected, data management includes cleaning and initial structuring, using the software knowledge and skills outlined above, and then transforming data into structured forms required for exploration, visualization, and analysis.
- **Data in Context – Capstone Experience** A capstone experience in which students consider scientific questions, collect and analyze data and communicate the results.

A possible path through the major is shown in Figure 1

# 5. Additional Considerations

1. Graduate Study
   Students interested in graduate study in mathematics, statistics or computer science may consider taking more advanced courses in theoretical foundations. The courses in mathematics for Data Science will not likely prepare a student for immediate acceptance into a PhD program in one of the three disciplines.
2. Articulation with community colleges
   Community colleges attempt to prepare students for many different purposes and institutions and, as a result, institutional change may be slow. In the meantime, given the existing course structure, students can prepare themselves to transfer to a college or university Data Science degree program.
   - Students can prepare by taking Calculus 1 and 2 as well as an Introduction to Computer Science course. Additional computer science courses, if offered, would be very helpful preparation. A few community colleges teach introductory statistics courses that emphasize data analysis and statistical thinking, and such courses should be required for transfer. More mathematical statistics courses that emphasize a rote, methods-based approach to statistics may not be an

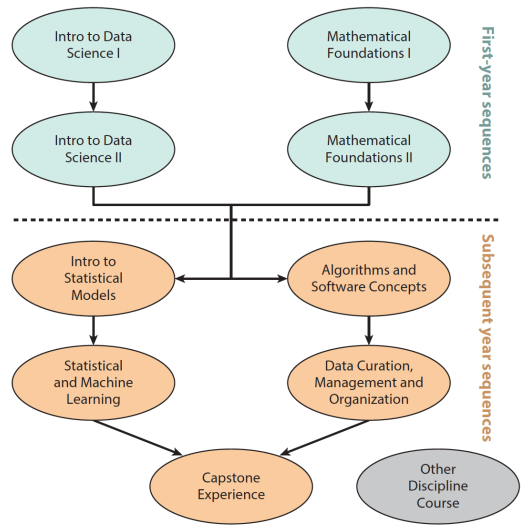

**Figure 1**
A Flow Chart Displaying a possible path through the major

       optimal preparation for data science.

- Not all community colleges will have the resources within a single department to develop a course such as the Data Science I and II courses proposed here. However, institutions should encourage collaboration between departments of mathematics and computer science in order to develop introductory statistics courses that (a) emphasize statistical thinking in the context of real and complex data sets (b) develop fundamental computational thinking through learning and using statistical software and (c) develop basic data handling skills, such as: creating new variables through transformations, uploading data with different delimiter types and different basic row/column structures, developing habits of reproducibility.

- The Statway (http://www.carnegiefoundation.org/resources/videos/introducing-statway/) and the New Mathways (http://www.utdanacenter.org/higher-education/new-mathways-project/) course sequences offered at some institutions may, depending on the local implementation, provide students with a strong grounding in statistical thinking and, if students learn to explore statistical concepts through simulations, also develop basic computational thinking.

3. Prerequisites and preparation in high school

    - As students exposed to Common Core standards for statistics enter college, some introductory material may need to be re-examined

    - To be prepared, a student must be calculus ready. *i.e.*. have a good precalculus course.

    - Basic matrix algebra (solving systems of linear equations) would be very helpful.

- Understanding what a scatterplot is, the basic concept of correlation and the line of best fit.
- Descriptive statistics: notions of center, spread, skew

4. Internship and applied experiences
   Internships and other applied experiences are a significant part of a Data Science program. Practical projects should be implemented often throughout the curriculum and provide the central experience of a capstone course.

## 6. Transitioning to a Data Science Major Using Typical Existing Courses

The curriculum we have outlined is founded on an integrated set of courses that span topics in three disciplines: mathematics, computer science and statistics. Many of these topics are covered in traditional courses found in those discipline. The highlighted courses show 10 courses that cover the bare minimum of the basics skills needed for Data Science. However, it is important to note that students will not be exposed to the richness of the interactions of these areas by taking only this set.

### 6.1. Courses in Mathematics

In order to best serve Data Science majors, math courses should emphasize connecting the mathematics to real world problems, especially data-driven problems

- **Calculus 1**
- **Calculus 2**
- Calculus 3
- **Linear Algebra**
- Probability Theory
- Discrete Math

### 6.2. Courses in Computer Science

The content in computational thinking courses is distributed across several traditional computer science courses, but not entirely contained in the three required courses listed.

- **Intro to Computer Science**
- **CS2: Data Structures/Algorithms**
- Computer Systems and Architecture
- Advanced Algorithms
- **Databases**
- Software Engineering

### 6.3. Courses in Statistics

Content in the Introduction to Statistics course should follow the revised Guidelines for Assessment and Instruction in Statistics Education for college courses (*http://www.amstat.org/education/gaise*)

- **Intro to Statistics**
- **Statistical Modeling/Regression**

- **Machine Learning/Data Mining**
- Theory of Statistics (requires Probability Theory)

### 6.4. Related Courses

- Introduction to [Partner Discipline]
- Intermediate course in Discipline
- **Capstone Course with Data Experience and Projects**
- Two courses in writing, preferably one in technical writing.
- Public Speaking
- Ethics

These ten highlighted courses cover the bare necessities of the material required for a Data Science major. We suspect many programs will want to add more courses to provide additional content and experience. For a more integrated major with 10 newly designed courses, please see the Appendix.

## 7. Summary and Next Steps

We hope that these guidelines can serve as a starting point for discussion for building new programs and transitioning existing programs.

In summary, the key points of our proposal involve:

> **SUMMARY POINTS**
>
> 1. Data Science is a fast evolving discipline centered on the acquisition, curation and analysis of data.
> 2. Courses from the traditional disciplines of mathematics, statistics and computer science provide the basic infrastructure for the major at present.
> 3. A redesign of the curriculum, integrating the elements of mathematical foundations and computational and statistical thinking at all levels will provide a rich and effective series of courses to prepare graduates for a career in Data Science.

We realize that the field is evolving rapidly, but hope that the basic areas we have outlined will be useful. During our discussions several issues arose that were outside the scope of our meeting, including the following:

> **FUTURE ISSUES**
>
> 1. **Faculty Development.** The courses outlined in the Appendix are clearly bold steps toward a new integrated program in Data Science. To be effective they will require many iterations. Resources for faculty including notes, examples, case studies and perhaps most important, new textbooks will be essential.
> 2. **Engagement with Two Year Colleges and High Schools.** The Data Science major will be attractive to many students coming from both high school and two year colleges. Interactions with these institutions will be crucial in order to coordinate courses and instruction to facilitate transfer to four-year institutions.

3. **Periodic Revision.** This is a first attempt at providing concrete guidelines for this emerging field. We realize that revisions will be necessary as the field continues to evolve and welcome feedback on these guidelines.

## ACKNOWLEDGMENTS

The authors would like to thank the Park City Mathematics Institute for supporting us in this effort. In addition, we would like to thank the National Science Foundation and the Institute for Advanced Study for supporting PCMI. The affiliations of the authors are as follows:

Richard D. De Veaux,[1] Mahesh Agarwal,[2] Maia Averett,[3] Benjamin S. Baumer,[4] Andrew Bray,[5] Thomas C. Bressoud,[6] Lance Bryant,[7] Lei Z. Cheng,[8] Amanda Francis,[9] Robert Gould,[10] Albert Y. Kim,[11] Matt Kretchmar,[12] Qin Lu,[13] Ann Moskol,[14] Deborah Nolan,[15] Roberto Pelayo,[16] Sean Raleigh,[17] Ricky J. Sethi,[18] Mutiara Sondjaja,[19] Neelesh Tiruviluamala,[20] Paul X. Uhlig,[21] Talitha M. Washington,[22] Curtis L. Wesley,[23] David White,[24] Ping Ye[25]

[1]Department of Mathematics and Statistics, Williams College, Williamstown, Massachusetts 01267

[2]Department of Mathematics and Statistics, University of Michigan, Dearborn, Michigan 48128-2406

[3]Department of Mathematics and Computer Science, Mills College, Oakland, California 94613

[4]Department of Statistical & Data Sciences, Smith College, Northampton, Massachusetts 01063

[5]Department of Mathematics, Reed College, Portland, Oregon 97202

[6]Department of Mathematics and Computer Science, Denison University, Granville, Ohio 43023

[7]Department of Mathematics, Shippensburg University, Shippensburg, Pennsylvania 17257

[8]Department of Mathematics, Olivet Nazarene University, Bourbonnais, Illinois 60914

[9]Department of Mathematics, Brigham Young University, Provo, Utah 84601

[10]Department of Statistics, University of California, Los Angeles, Los Angeles, California 90095-1554

[11]Department of Mathematics, Middlebury College, Middlebury, Vermont 05753

[12]Department of Mathematics and Computer Science, Denison University, Granville, Ohio 43023

[13]Department of Mathematics, Lafayette College, Easton, Pennsylvania 18042-1780

[14]Department of Mathematics and Computer Science, Rhode Island College, Providence, Rhode Island 02908

[15]Department of Statistics, University of California, Berkeley, California 94720

[16]Department of Mathematics, University of Hawaii, Hilo, Hawaii 96720-4091

[17]Department of Mathematics, Westminster College, Salt Lake City, Utah 84105

[18]Department of Computer Science, Fitchburg State University, Fitchburg, Massachusetts 01420

[19]Department of Mathematics, New York University, New York, New York 10012

[20]Department of Mathematics, University of Southern California, Los Angeles, California 90089

[21]Department of Mathematics, St. Mary's University, San Antonio, Texas 78228

[22]Department of Mathematics, Howard University, Washington, DC 20059

[23]Department of Mathematics, LeTourneau University, Longview, Texas 75602

[24]Department of Mathematics and Computer Science, Denison University, Granville, Ohio 43023

[25]Department of Mathematics, University of North Georgia, Oakwood, Georgia 30566

## LITERATURE CITED

MAA 2015. 2015 CUPM Curriculum Guide to Majors in the Mathematical Sciences *M*athematics Association of America (MAA) *http://www.maa.org/sites/default/files/pdf/CUPM/pdf/CUPMguide_print.pdf*

ACM 2013. Computer Science Curricula 2013: Curriculum Guidelines for Undergraduate Degree Programs in Computer Science (ACM) *https://www.acm.org/education/CS2013-final-report.pdf*

ASA 2015. Curriculum Guidelines for Undergraduate Programs in Statistical Science (ASA) *http://www.amstat.org/education/pdfs/guidelines2014-11-15.pdf*

Cassel & Topi. Strengthening Data Science Education Through Collaboration [Draft Title] Report on a Workshop on Data Science Education Funded by the National Science Foundation, Award #: DOE 1545135

ASA. Discovery with Data: Leveraging Statistics with Computer Science to Transform Science and Society", *www.amstat.org/policy/pdfs/BigDataStatisticsJune2014.pdf*

McKinsey.

Shron 2014. Shron,M. 2014 Thinking with Data: How to Turn Information into Insights. O'Reilly Media Inc. Sepastopol, CA.

MAA. CUPM Curriculum Guide 2004: Undergraduate programs and Courses in the Mathematical Sciencestinyurl.com/cupm2004. *http://www.maa.org/programs/faculty-and-departments/curriculum-department-guidelines-recommendations/cupm/cupm-guide-2004*

Horton & Hardin. Horton, N., Hardin, J. Teaching the Next Generation of Statistics Students to Think With Data: Special Issue on Statistics and the Undergraduate Curriculum. *The American Statistician* 69: 259-265.

Bryce *et al*. Bryce, G.R., Gould, R., Notz, W.I., Peck, R.L. ,2001, Curriculum Guidelines for Bachelor of Science Degrees in Statistical Science *American Statistician* 55(1)

2001. Breiman, L., 2001. Statistical Modeling: The Two Cultures, *Statistical Science* 16,3,199-231

2008. Wilkinson,L., (2008) The Future of Statistical Computing, *Technometrics* 50:4, 418-435

Wild & Pfannkuch. Wild, C.J., Pfannkuch, M. (1999), Statistical Thinking in Empirical Enquiry *International Statistical Review*, 67(3) 223-265.

Wing. Wing, J. (2006), Computational Thinking, *Communcations of the ACM* 49(3) 33-35.d

## 8. Appendix – Detailed Courses for a Proposed Data Science Major

1. **Intro to Data Science I**

    (a) Vision

    A complete 'alpha to omega' introduction to Data Science. Students will engage in the full data workflow including collaborative Data Science projects. This class is meant to be a high level introduction to the spectrum of Data Science topics, probably best taught in an iterative cycle from initial investigation and

data acquisition to the communication of final results
  (b) Learning Goals
      - Exploring and wrangling data
      - Writing basic functions and coding
      - Summarizing, visualizing, and analyzing data
      - Modeling and simulating deterministic and stochastic phenomena
      - Presenting the results of a complete project in written, oral, and graphical forms
  (c) Topics
      - What are data?
        - What forms do they take? What sorts of questions can they answer?
          * Description, prediction, inference
          * Implementation: Data should be non-trivially complex, but relatively clean. At this stage the data should be given to students and be sufficiently rich in terms of number of variables, size, multiple tables, etc.
      - Introduction to high level programming language + Integrated Development Environment (IDE) (R recommended)
      - Describing data: Exploratory Data Analysis (EDA) + Data Visualization
        - Summaries, aggregation, smoothing, distributions
      - Modeling + Stochastics (understand notions of uncertainty, simulations, random number generator, etc.)
        - Notion of mathematical model AKA function e.g. linear, exponential programming concepts
        - vectors, tables/data frames, variables
        - sequential programming / scripting
        - defining very simple functions, basic environment and scoping rules
        - conditional expressions and basic iteration
      - simulation w/wo data: : probabilistic and/or resampling based
      - Algorithms
        - breaking a complex problem down into small steps
        - concept of an algorithm as a 'recipe'
      - Project
        - Communication: Evaluate students on written, oral, and graphical forms
        - Ethics: *e.g.* copying code is plagiarism
        - Teamwork: working on a group project with version control (*e.g.* GitHub). Concepts of project and code management.
        - Motivate with application areas

2. **Intro to Data Science II**
   (a) Vision
       Exposure to different data types and sources, the process of data curation for the purpose of transforming them to a format suitable for analysis. Introduction

to the elementary notions in estimation, prediction and inference. We envision this class to be taught through case studies involving less-manicured data to enhance their computational and analytical abilities.

(b) Learning goals
- Interacting with a variety of data sources including relational databases
- Accessing data via different interfaces
- Building structure from a variety of data forms to enable analysis
- Formulating problems and bringing elementary concepts in estimation, prediction, and inference to bear
- Understanding how the data collection process influences the scope of inference

(c) Topics
- Kinds of data: *e.g.* static, spatial, temporal, text, media, ...
- Data sources: *e.g.* relational databases, web/API, streaming,
- Data collection: *e.g.* sampling, design (observational vs experimental) and its impact on visualization, modeling and generalizability of results
- Data cleaning/extraction: *e.g.* wrangling, regular expressions, SQL statements,
- Data analysis/modeling:
  - Question/problem formation along with EDA
  - Introduction to estimation and inference (testing and confidence intervals) including simulation and resampling
  - Scope of inference
  - Assessment and selection *e.g.* training and testing sets
- Project: A more comprehensive data-driven project including problem formulation, informed data wrangling and elementary analysis and conclusions including limitations to generalizability
  - Ability to collect data is enhanced
  - Better tools for analysis/scope of inference
  - Communicate, team, ethics, presentation

3. **Mathematics for Data Science I**

   (a) Vision

   Data science students need to use mathematical and statistical models, and these require mathematical foundations. This course introduces concepts from linear modeling and optimization, providing mathematical foundations as they are needed and motivated by applications. The focus is not on proof nor on excessive hand computations; instead, it is on employing and relating the mathematics to the real world ideas. Concepts are made concrete through numerical computation.

   (b) Learning Goals
   - Model real-world phenomena with functions (*e.g.* linear, polynomial, exponential)

- Optimize functions and interpret results
- Develop geometric intuition for linear modeling and optimization
- Employ tools for understanding local behavior of functions and models
- Learn tools from matrix algebra to analyze higher dimensional models
- Investigate sensitivity of models to small changes in the inputs
- Navigate the choice of model; understanding limitations of models
- Generate models for varied applications, using a diversity of tools

(c) Topics
- Introduction to modeling. Simple models via functions (*e.g.* from physics).
- Linear Algebra and Geometry: vectors, dot product, distance, projection, matrix algebra
- Multivariate thinking: (hyper)plane equations, systems of equations
- Sensitivity of models to changes in the input: slopes of lines, cross sections of planes, secant and tangent lines for curves
- Solving linear systems of equations:
  - Geometric approach to the determinant as a measurement of volume
  - Row operations, echelon form, and geometric interpretation
  - Degenerate matrices, non-existence of solutions, criterion for existence of solutions
- Derivatives:
  - Linear approximation
  - Optimization in the context of Data Science and regression
  - Gradient method of steepest descent
- Geometric approach to correlation via the dot product
- Matrix factorizations (if time allows)
  - LU and QR factorizations
  - Geometric interpretation of linear transformations
  - Eigenvalues and eigenvectors, diagonalization
  - Singular Value Decomposition
  - Principal Component Analysis

4. **Mathematics for Data Science II**

   (a) Vision

   Introduce probability theory and data-generating processes that lead to probability distributions. Build probability distributions from empirical data, and density functions from histograms. Topics of integration are introduced as motivated by probabilistic ideas and the transition from discrete data to continuous functions. The focus is not on proof nor on excessive hand computations; instead, ideas and concepts are made concrete through visualization and numerical computation.

   (b) Learning Goals
   - Understand and appreciate the role of randomness in data-generating processes.

- Apply properties and rules of probability to specific problems or questions.
- Use simulation at a basic level to generate data and to explore probability.
- Interpret area under a curve in a probabilistic context.
- Identify appropriate use of probability distributions, including ideas of expectation and variance.
- Extend univariate concepts to their multivariate analogues.

(c) Topics

- Randomness, random variables, empirical distributions
- Notions of probability: sample spaces and events, sets and counting, Venn diagrams
- Simulation as a constructive tool to explore probability; law of large numbers
- Conditional probability and independence: Bayes' formula
- Binomial, geometric, and normal distributions. Others taught "just in time" as needed in advanced courses.
- Integration
    - Motivate integrals via probability, expectation, and variance (discrete to continuous)
    - The fundamental theorem of calculus: relationship between probability distribution function and cumulative distribution function
    - Understanding basic numerical integration and the idea of error bounds
- Sampling distributions and the central limit theorem through simulation
- Joint probability distributions, multiple integrals, marginal distributions
- Series and connection to discrete probability distributions

5. **Algorithms and Software Concepts**

   (a) Vision

   Deepen students' understanding of computational problem solving begun in DS I and II. Continued use of a data-centered approach and employing scaffolding of problems and techniques, using procedural programming and exploring common problem solving strategies and assessing efficiency. Utilize statistical algorithms and data methods as a vehicle for learning the computer science concepts.

   (b) Learning Goals

   - Writing complete programs to solve problems
   - Learn a second programming language (*e.g.* Python, C++, Java)
   - Gain comfort with a new IDE (*e.g.* Eclipse)
   - Deepen programming language understanding, including concepts of variable scoping, data types, parameter passing, etc.
   - Understand implementations of basic statistical algorithms and their efficiency, from both abstract and lower level perspectives

   (c) Topics

   - Applying fundamental data structures useful in a Data Science context
       - lists (function application, aggregating statistical properties)

- stacks (encoding sequential data transformations)
- queues (simulating arrival processes)
- dictionaries (implementing graph representations, bag of words)
- matrices (implementing factorization algorithms, implementing group-by/summarize)
- trees (implementing recursively partitioned decision tree)
- efficiency of operations applied to the above data structures
- scalability of various data structures, including distribution beyond a single node
- Complexity of algorithms
  - Big-O notation, practical vs. theoretical efficiency
  - Brute force, divide-and-conquer
- Software development process
  - correctness, evaluation, debugging
  - code readability/documentation
  - project and code management – version control
- Comparing programming languages
  - functional programming centered view from DS I and II vs. imperative programming: the value of provable correctness, the ability to map an operation across an entire list, connection to system design, ethical implications
  - basics of object-oriented programming: software design using abstract data types
- High performance computing
  - algorithmic paradigms (*e.g.* map-reduce, online/streaming)
  - systems (*e.g.* Hadoop, Spark, NoSQL)

6. **Data Curation, Management, Organization**

   (a) Vision

   Deepen students' knowledge of data curation and management, building from the foundation of DS II by acquiring data across a spectrum of single and multiple sources and varying degrees of structure, incorporating real-world data throughout this spectrum. Gain understanding in how to transform and wrangle acquired data into forms amenable for analysis. Extend data management concepts and skills to accommodate big data.

   (b) Learning Goals
   - Acquire data from a spectrum of external systems, ranging from well-structured systems with defined schema to unstructured, requiring access through APIs and requiring scraping/munging, to large-scaled distributed storage systems
     - Acquire data from multiple sources. Be able to organize and unify them.
     - Be able to learn to learn – generalize to new APIs and new systems and be able to acquire data from them

- Understand the statistical issues inherent in the sources of data –bias, randomness or lack thereof – and the impact of those issues on analysis and generalization.
- Understand the structure of data needed to enable exploration, visualization, and analysis, and be able to take data not in that form and transform it appropriately
    - Advanced data cleaning and transformation
    - Blend tools from a variety of sources to combine and analyze data

(c) Topics

- Query languages and operations to specify and transform data (*e.g.* projection, selection, join, aggregate/group, summarize)
- Structured/schema based systems as users and acquirers of data
    - Relational (SQL) databases, APIs and programmatic access, indexing
    - XML and XPath, APIs for accessing and querying structured data contained therein
- Semistructured systems as users and acquirers of data
    - Access through APIs yielding JSON to be parsed and structured
- Unstructured systems in the acquisition and structuring of data
    - Web Scraping
    - Text/string parsing/processing to give structure
- Security and ethical considerations in relation to authenticating and authorizing access to data on remote systems
- Software development tools (*e.g.* github, version control)
- Large scale data systems
    - Paradigms for distributed data storage
    - Practical access to example systems (*e.g.* MongoDB, HBase, NoSQL systems)
        * Amazon Web Services (AWS) provides public data sets in Landsat, genomics, multimedia

7. **Intro to Statistical Models**

(a) Vision Introduces students to a framework for inference using regression models. The foundation is linear models, which is then compared to non-linear approaches. The course builds on important concepts introduced in the first year data science course that form the foundation of any statistical analysis. All the ideas are firmly grounded and inspired from real world data. A sound understanding of the ideas in this course will enable the student to understand the common structure underlying important algorithms and methods in Statistical and Machine Learning.

(b) Learning Goals

- Understand the mathematical framework for linear models and their applications.
- Apply regression models to real data sets for the purpose of scientific inference, understanding and communication.

- Introduce both theoretical and simulation approaches to uncertainty in models.

(c) Topics

- Review of hypothesis testing, confidence intervals, etc.
- Estimation *e.g.* likelihood principle, Bayes,
- Linear models
    - Regression theory *i.e.* least-squares: Introduction to estimation principles
    - Multiple regression
        * Transformations, model selection
        * Interactions, indicator variables, ANOVA
    - Generalized linear models *e.g.* logistic, etc.
- Alternatives to classical regression *e.g.* trees, smoothing/splines
- Introduction to model selection
    - Regularization, bias/variance tradeoff *e.g.* parsimony, AIC, BIC
    - Cross validation
    - Ridge regressions and penalized regression *e.g.* LASSO

8. **Statistical and Machine Learning**

(a) Vision

This is a course that blends the algorithmic perspective of machine learning in computer science and the predictive perspective of statistical thinking. It will focus on the common machine learning methods and their application to problems in various disciplines. The student will not only gain an understanding of the theoretical foundation of statistical learning, but the practical skills necessary for their successful application to new problems in science and industry.

(b) Learning Goals

- Gain experience with specific, widely used machine learning algorithms
- Understand the basic theoretical underpinnings of the methods and their limitations in inference
- Be able to apply the methods to real data, evaluate their performance and effectively communicate the results

(c) Topics

- Algorithms - what is the process doing (input/output), what can this solve.
    - Practical issues of model implementation (*e.g.* scalability, curse of dimensionality)
- Performance Metrics and Prediction
    - Loss functions
    - Model selection and assessment, *e.g.* cross-validation, performance measures, regularization methods
- Data transformations (make clear in the language what we mean here)
    - Dimension reduction, PCA, feature extraction

- Smoothing and aggregating: *e.g.* kernel methods, spatial and/or temporal smoothing, model averaging.
- Supervised Learning
  - Regression: linear models, regression trees
  - Classification: classification trees, logistic reg, separating hyperplanes, k-NN
- Unsupervised Learning
  - Clustering: k-means, hierarchical
- Ensemble methods *e.g.* boosting and bagging.

(d) Project (Common Task Framework)

9. **Capstone Course**

   (a) Vision

   The capstone course provides students with a comprehensive learning experience that integrates knowledge from previous courses. The student will make connections among ideas and experiences gained from all three core disciplines and apply them to another domain. In the course students will synthesize what they have learned and apply that knowledge to new, complex situations. Student should engage in the entire process of solving a real-world data science project: from collecting and processing actual data, applying suitable and appropriate analytic methods to the problem and communicating the results in a clear and comprehensible way. The capstone should emphasize a good understanding of the foundational knowledge of the core disciplines and the domain area, and prepares the student for future professional endeavors.

   (b) Learning Goals

   - Ability to handle a problem in data science from the initial problem formulation through a proposed solution and then communicating the results of that effort. The student will demonstrate proficiency in the entire process of collecting, preparing and analyzing real-world data to provide a solution to the initial problem.
   - Learn how to work in teams by working with at least one other student on their project.
   - Present results and code which are reproducible and ethical.
   - Present a report (both oral and written) on the project emphasizing
     i. Motivation and problem definition, existing approaches to the problem
     ii. proposed solution
     iii. Results, conclusions, limits to the present analysis, and directions for future work.

   (c) Suggested areas and topics for capstone projects:

   - **Application project** A project mainly focused on an application of interest to the student. The student should explore how best to apply learning algorithms to solve the problem of interest, and be able to evaluate the performance and the limitations of the model.

- **Algorithmic/Computational project** Motivated either by existing algorithms and the literature pertaining to them, or from an application area, a student may propose a new learning algorithm, a novel variant of an existing algorithm, or a novel application of an existing algorithm.
- **Theoretical project** For the more ambitious student, analyzing and possibly proving some properties of a new or an existing learning algorithm.